\documentclass[english,osajnl,jcp,preprint,groupedaddress,noshowpacs,showkeys,longbibliography,noeprint,nofootinbib,bibtex,ruled,vlined,english,natural,table,usenames,dvipsnames]{revtex4-1}
\usepackage{lmodern}
\usepackage{lmodern}
\usepackage[T1]{fontenc}
\usepackage[utf8]{inputenc}
\usepackage{color}
\usepackage{babel}
\usepackage{array}
\usepackage{booktabs}
\usepackage{mathtools}
\usepackage{amsmath}
\usepackage{amssymb}
\usepackage{graphicx}
\PassOptionsToPackage{version=3}{mhchem}
\usepackage{mhchem}
\usepackage{microtype}
\usepackage[unicode=true,
 bookmarks=false,
 breaklinks=false,pdfborder={0 0 1},backref=section,colorlinks=true]
 {hyperref}
\hypersetup{
 linkcolor=BrickRed,urlcolor=blue!50!black,citecolor=blue!50!black}

\makeatletter

\providecommand{\tabularnewline}{\\}

\usepackage{cmap}% make ligatures etc. copy- and searchable
\usepackage{mleftright}% fix spacing of \left( and right)

\usepackage{grffile}\graphicspath{{./}}
\newlength{\onecolwidth}
\newlength{\twocolwidth}
\usepackage[ruled,vlined]{algorithm2e}

\usepackage{url}
\usepackage{xcolor}

\makeatother

\begin{document}

\title{Reversible Interacting-Particle Reaction Dynamics}

\author{Christoph Fröhner}
\email{christoph.froehner@fu-berlin.de}

\affiliation{Freie Universität Berlin, Fachbereich Mathematik und Informatik,
Arnimallee 6, 14195 Berlin, Germany }

\author{Frank Noé}
\email[corresponding author: ]{frank.noe@fu-berlin.de}

\affiliation{Freie Universität Berlin, Fachbereich Mathematik und Informatik,
Arnimallee 6, 14195 Berlin, Germany }

\date{\today}
\begin{abstract}
Interacting-Particle Reaction Dynamics (iPRD) simulates the spatiotemporal
evolution of particles that experience interaction forces and can
react with one another. The combination of interaction forces and
reactions enable a wide range of complex reactive systems in biology
and chemistry, but give rise to new questions such as how to evolve
the dynamical equations in a computationally efficient and statistically
correct manner. Here we consider reversible reactions such as $\mathrm{A}+\mathrm{B}\rightleftarrows\mathrm{C}$
with interacting particles and derive expressions for the microscopic
iPRD simulation parameters such that desired values for the equilibrium
constant and the dissociation rate are obtained in the dilute limit.
We then introduce a Monte-Carlo algorithm that ensures detailed balance
in the iPRD time-evolution (iPRD-DB). iPRD-DB guarantees the correct
thermodynamics at all concentrations and maintains the desired kinetics
in the dilute limit, where chemical rates are well-defined and kinetic
measurement experiments usually operate. We show that in dense particle
systems, the incorporation of detailed balance is essential to obtain
physically realistic solutions. iPRD-DB is implemented in ReaDDy~2
(\href{https://readdy.github.io}{https://readdy.github.io}). 
\end{abstract}

\maketitle

\section{Introduction}

Particle based reaction diffusion (PBRD) dynamics is a detailed model
for simulating the spatiotemporal evolution of reactive particles~\citep{Erban2009,VanZon2005,Hoffmann2014,Andrews2017}.
Resolving the trajectories of every reactive particle is important
in applications where the reactants cannot be assumed to be spatially
well-mixed~\citep{Agarwal2016,AlbrechtEtAl_JCP16_Nanoscopic} or
always sufficiently abundant to be described by a continuous concentration~\citep{Elowitz2002,Bhalla2004}
-- e.g., consider many cases of cellular signalling and reactions
in nontrivial architectures~\citep{Doherty2009,Pollard2003,Sadiq2016}.
A common implementation of PBRD is to propagate particle positions
with overdamped Langevin dynamics (Brownian motion) in discrete time
steps, and execute discrete reaction events such as $\mathrm{A}+\mathrm{B}\rightarrow\mathrm{C}$
with a certain probability when two reactive particles $\mathrm{A}$
and $\mathrm{B}$ are close in space. When the system is sufficiently
dilute, such simulations can be sped up by exploiting solutions of
the one- or two-particle diffusion equation~\citep{VanZon2005,van2005simulating,Takahashi2010,Opplestrup2006,Donev2010}.

A recent extension of PBRD is the interacting-Particle Reaction Dynamics
(iPRD) method~\citep{Schoneberg2013,Schoneberg2014,Biedermann2015},
in which particles are additionally subject to interaction forces.
Alternatively, iPRD could be characterized as a form of coarse-grained
Molecular Dynamics (MD) simulation with reactions between particles.
Particle interaction forces are useful to model order and structure
on mesoscopic lengthscales, such as the space-exclusion in dense particle
systems~\citep{Schoneberg2013,Hofling2013}, the restriction of diffusing
particles to arbitrarily-shaped membranes~\citep{Schoneberg2013,Gunkel2015,Schoneberg2014a},
the large-scale structure of polymers~\citep{readdy2} and membranes~\citep{Sadeghi2018},
and the clustering of attractive proteins~\citep{Ullrich2015}. The
combination of interaction forces and reactions allow an even wider
range of complex reactive systems in biology and chemistry to be modeled,
such as the dynamics of phototransduction that involve protein diffusion
in particle-dense photoreceptor membranes~\citep{Schoneberg2014a},
the effect of transmembrane protein oligomers on these dynamics~\citep{Gunkel2015},
the recruitment of proteins to endosomes~\citep{Posor2013,Schoneberg2017},
and the assembly, diffusion, and dissociation of polymers~\citep{readdy2}.
The idea of combining PBRD with particle interaction forces is also
found in MD-GFRD~\citep{Vijaykumar2015,Vijaykumar2017}, where the
close particle interactions are simulated by MD and the reaction-diffusion
model is used to derive an efficient way to propagate particles while
they are not-interacting. In contrast, in iPRD particle interactions
and reactions occur simultaneously, with the idea that reaction events
are a suitable way to coarse-grain complicated events such as protein-protein
binding, whose kinetics might be obtained from Markov State Models
of all-atom MD simulations~\citep{Dibak2018}. MD-GFRD simulations
can be used to speed up iPRD simulations when the system is sufficiently
dilute~\citep{Vijaykumar2015,Vijaykumar2017,Sbailo2017}, and with
free-propagator reweighting, this speedup can also be obtained in
the regime where particles are interacting~\citep{Johnson2014}. 

An open question is: What is the statistically correct way to model
the dynamical evolution of simultaneously interacting and reacting
particles? Specifically, we consider reversible reactions, such as
$\mathrm{A}+\mathrm{B}\rightleftarrows\mathrm{C}$, as they are found
in nature, but also in technological applications. Examples include
reversible protein-drug binding~\citep{Scott2016,Paul2017}, reversible
protein-protein association that can now be simulated at atomistic
detail~\citep{Plattner2017}, and metal ion deposition to / removal
from electrodes in batteries that are driving charging and discharging~\citep{Armand2008,Boyanov2006}.
To derive a statistically correct simulation scheme of $\mathrm{A}+\mathrm{B}\rightleftarrows\mathrm{C}$
via iPRD, we need to answer the following questions:
\begin{enumerate}
\item Which bimolecular reaction scheme should be used, i.e. under which
conditions will two particles $\mathrm{A}$ and $\mathrm{B}$ fuse
into a $\mathrm{C}$ particle?
\item How do we choose the microscopic parameters of this reaction scheme
such that the iPRD simulation samples the macroscopic kinetic quantities
that have been obtained from experiments or more detailed MD simulations?
\item When executing $\mathrm{A}+\mathrm{B}\rightarrow\mathrm{C}$ or $\mathrm{C}\rightarrow\mathrm{A}+\mathrm{B}$,
where should the product particles be placed, such that the simulation
obeys detailed balance?
\end{enumerate}
The answers to these three questions are coupled. 

Question 1: For the sake of analytical computations, the best-studied
reaction scheme is the Smoluchowski model where diffusing particles
react instantly when they establish contact, defined by a reaction
distance $R$~\citep{Smoluchowski1916}. The Collins-Kimball model~\citep{Collins1949}
reduces the probability of reacting upon contact to a finite value
$\le1$. Reversible reactions in the Collins-Kimball model are discussed
in~\citep{DelRazo2016}, for interacting particles of isolated pairs
an analytical description is found in~\citep{Agmon1990}. In iPRD
simulations we instead use the Doi model~\citep{Teramoto1967,Doi1976}:
\[
\mathrm{A}+\mathrm{B}\rightleftarrows\mathrm{A}\mathrm{B}\rightleftarrows\mathrm{C}
\]
Here two particles $\mathrm{A}$ and $\mathrm{B}$ form a reactive
complex $\mathrm{A}\mathrm{B}$ when their distance is less or equal
to $R$. This process is simulated by the dynamical model that propagates
particles (e.g. overdamped Langevin equation). Whenever a reactive
complex $\mathrm{A}\mathrm{B}$ exists, it can decay to a $\mathrm{C}$
particle with a microscopic rate constant $\lambda$. The reverse
process happens with a microscopic rate constant $k_{\mathrm{off}}$.
The Doi model is well compatible with a finite-time-stepping simulation
scheme, where the formation of $\mathrm{A}\mathrm{B}$ can be easily
checked in every time-step as part of the particle neighbor list update.

Question 2: When using the Doi model, how should the parameters in
this model be chosen? The dissociation rate constant $k_{\mathrm{off}}$
can be directly obtained from kinetic experiments or all-atom MD simulations
with accelerated sampling methods~\citep{Plattner2017,Paul2017,Du2014}.
For the Doi model where $\mathrm{A}$ and $\mathrm{B}$ encounter
from a long distance via normal diffusion without interaction forces,
the association parameters $R$ and $\lambda$ can be computed from
an equation derived in~\citep{Doi1975,Erban2009}. When $\mathrm{A}$
and $\mathrm{B}$ interact, such a result can still be obtained numerically~\citep{DibakFroehner2018}.
In Sec.~\ref{sec:bi-mol-equilibrium} we develop a theory for the
$\mathrm{A}+\mathrm{B}\rightleftarrows\mathrm{C}$ reaction of an
isolated pair, that is independent of the diffusion coefficient $D$.
This enables to choose $\lambda$ for given dissociation rate constant
$k_{\mathrm{off}}$, interaction radius $R$, and $\mathrm{A}-\mathrm{B}$
interaction potential such that the iPRD simulation will produce a
desired equilibrium constant and association rate constant at low
particle concentrations, as they are typically found in experiments
measuring these constants.

Question 3: Time-reversible processes evolving in thermodynamic equilibrium
obey detailed balance~\citep{VanKampen1992}. For example, consider
that we have system with one particle $\mathrm{A}$ and $\mathrm{B}$
each at positions $\mathbf{x}_{A},\mathbf{x}_{B}$ and we perform
the forward reaction to a system with one particle $\mathrm{C}$ at
position $\mathbf{x}_{C}$. Detailed balance implies that the equilibrium
probability of being in the $\mathrm{A},\mathrm{B}$ system at $\mathbf{x}_{A},\mathbf{x}_{B}$
times the forward reaction rate must be equal to the equilibrium probability
of being in the $\mathrm{C}$ system at $\mathbf{x}_{C}$ times the
backward reaction rate, and this must be true for all system configurations.
Vice versa, enforcing detailed balance is a technically convenient
way to automatically achieve a desired equilibrium distribution. It
implies a relationship between forward and backward reaction rates
and also that the reaction scheme that allows for a forward reaction
$\mathbf{x}_{A},\mathbf{x}_{B}\rightarrow\mathbf{x}_{C}$ must also
allow for the reverse reaction, and vice versa. For non-interacting
PBRD, a detailed balance scheme was first introduced in~\citep{Morelli2008}.
Other schemes have been developed more recently~\citep{Klein2014,Donev2018}.
In Sec.~\ref{sec:db-general}, we develop a general detailed-balance
scheme for iPRD (iPRD-DB). The scheme includes a Metropolis-Hastings~\citep{Metropolis1953,Hastings1970}
acceptance step that ensures the resulting dynamics fulfill detailed
balance for abitrary configurations of interacting particles. In the
dilute limit (one $\mathrm{A}$ and $\mathrm{B}$ particle pair reacting
to a single $\mathrm{C}$ particle and back), the proposal steps are
designed such that they are always accepted and the desired equilibrium
association and dissociation rate constants are obtained. When the
so-parametrized particles enter a dense phase, the kinetics and equilibria
will naturally change, but do so in a physically realistic manner.
In particular, we show that in a dense particle system where the reaction
$\mathrm{A}+\mathrm{B}\rightleftarrows\mathrm{C}$ involves a change
in effective particle volume, the iPRD-DB scheme leads to a solution
that is consistent with Le Chatelier's principle, while a regular
Doi scheme that ignores detailed balance produces unphysical solutions.

The implementation of the iPRD-DB scheme is included in the ReaDDy~2
software package~\citep{readdy2}. 

\section{Bimolecular reaction in equilibrium}

\label{sec:bi-mol-equilibrium}
\begin{figure}
\centering\includegraphics[width=3.4in]{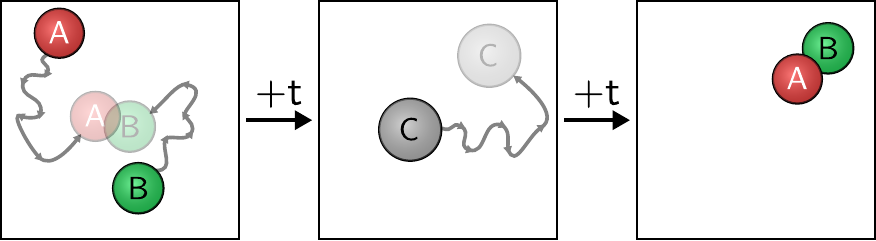}

\caption{Schematic time evolution of a reaction-diffusion system of an isolated
pair subject to the reaction $\mathrm{A}+\mathrm{B}\rightleftarrows\mathrm{C}$
with the Doi model. Particles $\mathrm{A}$ and $\mathrm{B}$ diffuse
and can form a complex particle $\mathrm{C}$ when they are closer
than a certain reaction radius, here depicted as the sum of the radii
of the two particles. The complex particle $\mathrm{C}$ diffuses
as well and can dissociate into $\mathrm{A}$ and $\mathrm{B}$ again.}
\label{fig:scheme}
\end{figure}

We consider a system of molecules with
three species, in which molecules $\mathrm{A}$ and $\mathrm{B}$
reversibly form a complex $\mathrm{C}$. We want to simulate particle
dynamics involving such reactions with iPRD, where particles interact
with a potential when they are close, and a certain microscopic reaction
scheme is employed, see Fig.~\ref{fig:scheme}. This section answers
the question how the microscopic parameters of this reaction scheme
need to be chosen such that the equilibrium constant and the dissociation
rate measurable in a bulk experiments will be reproduced. This result
will be used in the next section as part of designing a scheme obeying
detailed balance.

\subsection{Macroscopic rate model}

The macroscopic reaction dynamics is described by the scheme
\begin{equation}
\mathrm{A}+\mathrm{B}\underset{k_{\mathrm{off}}}{\overset{k_{\mathrm{on}}}{\rightleftarrows}}\mathrm{C},\label{eq:bimol-reaction}
\end{equation}
where $k_{\mathrm{on}}$ is a macroscopic bimolecular association
rate constant, measured in units of per time and per concentration,
while $k_{\mathrm{off}}$ is the dissociation rate constant, measured
in units of per time. These are related to the macroscopic dissociation
constant $K_{d}$, measured in units of concentration: 
\begin{equation}
K_{d}=\frac{k_{\mathrm{off}}}{k_{\mathrm{on}}}.\label{eq:dissociation-constant}
\end{equation}
We assume that both the association- and the dissociation process
obey a linear rate law~\citep{Atkins2006}, according to the law
of mass action (LMA). We define the effective association rate $K_{\mathrm{on}}$
\begin{equation}
K_{\mathrm{on}}=k_{\mathrm{on}}V^{-1},\label{eq:macroscopic-effective-on-rate}
\end{equation}
which is the frequency of association per $\mathrm{AB}$ complex.
Likewise we define the effective dissociation rate $K_{\mathrm{off}}$
\begin{equation}
K_{\mathrm{off}}=k_{\mathrm{off}},\label{eq:macroscopic-effective-off-rate}
\end{equation}
which is the frequency of dissociation per $\mathrm{C}$ molecule.
We denote $\pi_{i}$ as the stationary probability of state $i$.
The ratio of stationary probabilities $\pi_{AB}/\pi_{C}$ is given
by the ratio of effective rates in equilibrium, where the number of
association events per time is equal to the number of dissociation
events per time 
\begin{equation}
\frac{\pi_{AB}}{\pi_{C}}=\frac{K_{\mathrm{off}}}{K_{\mathrm{on}}}=\frac{[A]_{\mathrm{eq}}[B]_{\mathrm{eq}}}{[C]_{\mathrm{eq}}}V=K_{d}V.\label{eq:equilibrium-constant}
\end{equation}

\subsection{Microscopic distribution}

\label{sec:microscopic-distribution}For the following we will assume
that there is only either one pair of $\mathrm{A}$ and $\mathrm{B}$
particles or one $\mathrm{C}$ particle which live inside the volume
$V$. The vectors $\mathbf{x}\in\mathbb{R}^{9}$, contain the euclidean
positions for three particles. Individual positions are denoted by
$\mathbf{x}_{a}$, $\mathbf{x}_{b},$ and $\mathbf{x}_{c}$ for particles
$\mathrm{A}$, $\mathrm{B}$, and $\mathrm{C}$ respectively. Additionally
there is a phase $i\in\{AB,C\}$, where $AB$ is the dissociated phase
and $C$ is the associated phase. The joint distribution for states
$x_{i}=(\mathbf{x},i)\in\mathbb{R}^{9}\times\{AB,C\}$ of finding
the system in phase $i$ and particle positions $\mathbf{x}$ is 
\begin{equation}
p(x_{i})=\left\{ \begin{array}{ll}
\pi_{AB}\,p_{AB}(\mathbf{x}) & \text{for }i=AB\\
\pi_{C}\,p_{C}(\mathbf{x}) & \text{for }i=C
\end{array}\right.\label{eq:micro-distr}
\end{equation}
Note that in phase $AB$ there is still a position for the $\mathrm{C}$
particle, such that the dimension of the microscopic phase space is
equal for both phases. The same occurs for the positions of $\mathrm{A}$
and $\mathrm{B}$ in the phase $C$. In both cases, the residual variables
have no effect. In phase space integrals these will be accounted for
by a volume factor. Hence all phase space integrals use the measure
$\mathrm{d}\mathbf{x}=\mathrm{d}\mathbf{x}_{a}\mathrm{d}\mathbf{x}_{b}\mathrm{d}\mathbf{x}_{c}$,
where each $\mathrm{d}\mathbf{x}_{j}$ has units of volume. Introducing
a Fock space for treatment of changing number of particles is circumvented
by considering at most three particles - the isolated pair and the
complex - and having the non existing particles contribute a constant
factor to the partition function.

In phase $AB$ the two particles $\mathrm{A}$ and $\mathrm{B}$ are
subject to an interaction potential $U(\mathbf{x})=U(|\mathbf{x}_{b}-\mathbf{x}_{a}|)=U(r)$
depending only on the distance $r=|\mathbf{x}_{b}-\mathbf{x}_{a}|$
of $\mathrm{A}$ and $\mathrm{B}$. The potential is cut off at $R_{\mathrm{int}}$,
i.e. $U(r)=0$, if $r>R_{\mathrm{int}}.$ The stationary distribution
of positions $\mathbf{x}$ in phase $AB$ is 
\[
p_{AB}(\mathbf{x})=Z_{AB}^{-1}\exp(-\beta U(r))\quad\text{with\ensuremath{\quad}}r=|\mathbf{x}_{b}-\mathbf{x}_{a}|
\]
where $\beta^{-1}=k_{B}T$ is the thermal energy of the system which
is coupled to a heat bath with temperature $T$and the normalization
constant can be computed as follows (see Appendix A), 
\begin{align}
Z_{AB} & =V^{2}(V-V_{\mathrm{ex}})\label{eq:partition-function-ab}\\
V_{\mathrm{ex}} & =V_{\mathrm{int}}-V_{\mathrm{int}}^{\mathrm{eff}}\label{eq:v-ex}\\
V_{\mathrm{int}} & =\frac{4}{3}\pi R_{\mathrm{int}}^{3}\\
V_{\mathrm{int}}^{\mathrm{eff}} & =\int_{0}^{R_{\mathrm{int}}}e^{-\beta U(r)}4\pi r^{2}\mathrm{d}r,\label{eq:v-int-eff}
\end{align}
where $V_{\mathrm{int}}$ is the interaction volume of the reactive
particles, $V_{\mathrm{int}}^{\mathrm{eff}}$ the effective accessible
volume due to particle interaction and $V_{\mathrm{ex}}$ is the reduction
of the accessible volume.

In phase $C$ the stationary distribution of positions $\mathbf{y}$
is 
\[
p_{C}(\mathbf{y})=Z_{C}^{-1}
\]
with the partition function 
\[
Z_{C}=\int\mathrm{d}\mathbf{y}=\iiint\mathrm{d}\mathbf{y}_{a}\mathrm{d}\mathbf{y}_{b}\mathrm{d}\mathbf{y}_{c}=V^{3}.
\]

\subsection{Doi reaction model}

The microscopic reaction model is defined by the association rate
function $\lambda^{+}(\mathbf{x})$ and the dissociation rate function
$\lambda^{-}(\mathbf{y})$. The former describes the probability per
unit time with which two particles $\mathrm{A}$ and $\mathrm{B}$
can react when the system is in phase $AB$ and depends on positions
$\mathbf{x}$. The latter describes the probability per unit time
with which a $\mathrm{C}$ particle dissociates into $\mathrm{A}$
and $\mathrm{B}$ when the system is in phase $C$. We assume that
$\lambda^{+}(\mathbf{x})$ is radially symmetric, i.e. it only depends
on $r=|\mathbf{x}_{b}-\mathbf{x}_{a}|$. Any microscopic reaction
model, described by $\lambda^{+}(\mathbf{x})$ will result in an effective
association rate $K_{\mathrm{on}}^{\mathrm{micro}}$ which reads 
\begin{equation}
\begin{aligned}K_{\mathrm{on}}^{\mathrm{micro}} & =\int\lambda^{+}(\mathbf{x})p_{AB}(\mathbf{x})\mathrm{d}\mathbf{x}\end{aligned}
\label{eq:eff-on-rate-micro-general}
\end{equation}
For $\lambda^{+}(\mathbf{x})$ and $\lambda^{-}(\mathbf{y})$ we use
the Doi reaction model as depicted in Fig.~\ref{fig:scheme}, i.e.
the microscopic association reaction rate function is a constant $\lambda_{\mathrm{on}}$,
when particles $\mathrm{A}$ and $\mathrm{B}$ are closer than the
reaction radius $R_{\mathrm{reac}}$
\begin{equation}
\lambda^{+}(\mathbf{x})=\lambda_{\mathrm{on}}\,\chi_{\mathrm{reac}}(r)\quad\text{with\ensuremath{\quad}}r=|\mathbf{x}_{b}-\mathbf{x}_{a}|,\label{eq:lambda-on-doi}
\end{equation}
where $\chi_{\mathrm{reac}}(r)$ indicates that $\mathrm{A}$ and
$\mathrm{B}$ are within reactive distance
\begin{equation}
\chi_{\mathrm{reac}}(r)=\left\{ \begin{array}{rl}
1, & \mathrm{if}\,r<R_{\mathrm{reac}}\\
0, & \mathrm{otherwise.}
\end{array}\right.
\end{equation}
The microscopic dissociation rate function is constant and chosen
equal to the macroscopic dissociation rate constant 
\begin{equation}
\lambda^{-}(\mathbf{y})=k_{\mathrm{off}}.\label{eq:lambda-off-doi}
\end{equation}
We evaluate the effective microscopic association rate~(\ref{eq:eff-on-rate-micro-general})
for the Doi reaction model~(\ref{eq:lambda-on-doi}) and obtain 
\begin{align}
K_{\mathrm{on}}^{\mathrm{micro}} & =\lambda_{\mathrm{on}}Z_{AB}^{-1}V^{2}\int_{0}^{R_{\mathrm{reac}}}e^{-\beta U(r)}4\pi r^{2}\mathrm{d}r\nonumber \\
 & =\lambda_{\mathrm{on}}\frac{V_{\mathrm{reac}}^{\mathrm{eff}}}{V-V_{\mathrm{ex}}}\label{eq:eff-on-rate-micro-doi}
\end{align}
where the effective reaction volume $V_{\mathrm{reac}}^{\mathrm{eff}}$
takes a similar form as the effective interaction volume , but with
another radius $R_{\mathrm{reac}}$ 
\begin{equation}
V_{\mathrm{reac}}^{\mathrm{eff}}=\int_{0}^{R_{\mathrm{reac}}}e^{-\beta U(r)}4\pi r^{2}\mathrm{d}r.\label{eq:v-reac-eff-doi}
\end{equation}

\subsection{Computing the microscopic association rate constant that reproduces
the macroscopic equilibrium\label{subsec:computing-microscopic}}

For the following we will assume a given dissociation constant $K_{d}$
and a given dissociation rate constant $k_{\mathrm{off}}$. Using
Eqs.~(\ref{eq:macroscopic-effective-on-rate},~\ref{eq:dissociation-constant})
we state the effective association rate according to the law of mass
action 
\begin{equation}
K_{\mathrm{on}}=\frac{k_{\mathrm{off}}}{K_{d}V}.\label{eq:eff-on-rate-macro}
\end{equation}
We require that the micro- and macroscopic effective rates match 
\begin{equation}
K_{\mathrm{on}}^{\mathrm{micro}}\overset{!}{=}K_{\mathrm{on}}\label{eq:couple-on-rates}
\end{equation}
and find the restrictions on the microscopic reaction model. This
results in a choice %\footnote{Note that this is only one particular choice. The proposed detailed balance method does not depend on this result.}
for the microscopic association rate constant $\lambda_{\mathrm{on}}$,
that will yield the desired equilibrium as in Eq.~(\ref{eq:equilibrium-constant}).
We will call this specific value $\tilde{\lambda}_{\mathrm{on}}$
\begin{equation}
\tilde{\lambda}_{\mathrm{on}}=\frac{k_{\mathrm{off}}}{K_{d}V}\frac{V-V_{\mathrm{ex}}}{V_{\mathrm{reac}}^{\mathrm{eff}}}.\label{eq:choice-on-rate}
\end{equation}
The relation of this expression to other diffusion influenced rate
calculations is discussed in Appendix B. 

\section{Interacting-Particle Reaction Dynamics with Detailed Balance}

\label{sec:db-general} Transition rates $k^{+}$ and $k^{-}$ of
association ($+$) and dissociation ($-$) respectively between states
$x_{AB}$ and $y_{C}$, with stationary probability distributions
$p$ defined in Eq.~(\ref{eq:micro-distr}) shall obey detailed balance
\begin{equation}
p(x_{AB})k^{+}(\mathbf{y}|\mathbf{x})=p(y_{C})k^{-}(\mathbf{x}|\mathbf{y}).\label{eq:db}
\end{equation}
We split the transition rates $k$ into proposal rate and acceptance
probability 
\begin{equation}
\begin{array}{rl}
k^{+}(\mathbf{y}|\mathbf{x}) & =\lambda^{+}(\mathbf{x})q^{+}(\mathbf{y}|\mathbf{x})\,\,\alpha^{+}(\mathbf{y}|\mathbf{x})\,\quad\text{association}\\
k^{-}(\mathbf{x}|\mathbf{y}) & =\underbrace{\lambda^{-}(\mathbf{y})q^{-}(\mathbf{x}|\mathbf{y})}_{\text{proposal}}\underbrace{\alpha^{-}(\mathbf{x}|\mathbf{y})}_{\text{acceptance}}\quad\text{dissociation}
\end{array}\label{eq:split}
\end{equation}
where $\lambda^{+}(\mathbf{x})$ is the absolute rate of proposing
a transition $\mathrm{A}+\mathrm{B}\rightarrow\mathrm{C}$ when in
particle configuration $\mathbf{x}$. $q^{+}(\mathbf{y}|\mathbf{x})$
is the normalized density to propose the positions $\mathbf{y}$,
given that the positions were $\mathbf{x}$ . $\alpha^{+}(\mathbf{y}|\mathbf{x})$
is the absolute probability of accepting the proposed positions. Similarly
$\lambda^{-}(\mathbf{y})$ is the absolute rate of proposing a transition
$\mathrm{C}\rightarrow\mathrm{A}+\mathrm{B}$, $q^{-}(\mathbf{x}|\mathbf{y})$
is the according proposal density and $\alpha^{-}(\mathbf{x}|\mathbf{y})$
the absolute probability of accepting the proposal. All $q$ and $\alpha$
satisfy 
\[
\int q^{i}(\mathbf{y}|\mathbf{x})\mathrm{d}\mathbf{y}=1\quad\text{and}\quad\alpha^{i}(\mathbf{y}|\mathbf{x})\leq1\quad\text{for }i\in\{+,-\}
\]

\subsection{Derive the backward proposal from the forward proposal}

We assume the association proposal density $q^{+}$ as given, and
want to derive the dissociation proposal density $q^{-}$ and both
$\alpha^{+}$ and $\alpha^{-}$ subject to detailed balance. Therefore
we include all terms that depend on the particle positions into the
reverse proposal density $q^{-}$, such that most terms in Eq.~(\ref{eq:db})
cancel and acceptances $\alpha^{+}$ and $\alpha^{-}$ become independent
of the particle positions of the dissociated phase. The reverse proposal
density reads 
\begin{equation}
q^{-}(\mathbf{x}|\mathbf{y})=Q(\mathbf{y})^{-1}q^{+}(\mathbf{y}|\mathbf{x})\frac{\lambda^{+}(\mathbf{x})}{\lambda^{-}(\mathbf{y})}\frac{p_{AB}(\mathbf{x})}{p_{C}(\mathbf{y})}\label{eq:proposal-reverse}
\end{equation}
with the normalization function $Q(\mathbf{y})$ such that 
\begin{equation}
Q(\mathbf{y})=\frac{1}{\lambda^{-}(\mathbf{y})p_{C}(\mathbf{y})}\int q^{+}(\mathbf{y}|\mathbf{x})\lambda^{+}(\mathbf{x})p_{AB}(\mathbf{x})\mathrm{d}\mathbf{x}\label{eq:q-norm-general}
\end{equation}
Note that $Q$ must depend on $\mathbf{y}$ to fulfil the normalization
$\forall\mathbf{y}$ (in the Doi model it will reduce to a constant).
Inserting Eqs.~(\ref{eq:proposal-reverse},~\ref{eq:split}) into
Eq.~(\ref{eq:db}), the detailed balance condition reduces to 
\begin{equation}
\frac{\alpha^{+}(\mathbf{y}|\mathbf{x})}{\alpha^{-}(\mathbf{x}|\mathbf{y})}=\frac{\pi_{C}}{\pi_{AB}}\frac{1}{Q(\mathbf{y})}\label{eq:db-reduced}
\end{equation}
Reminding that $\alpha\leq1$ naturally leads to using the Metropolis-Hastings~\citep{Hastings1970,Metropolis1953}
acceptance function 
\begin{equation}
\begin{aligned}\alpha^{+}(\mathbf{y}|\mathbf{x}) & =\min\left\{ 1,\frac{\pi_{C}}{\pi_{AB}\,Q(\mathbf{y})}\right\} \\
\alpha^{-}(\mathbf{x}|\mathbf{y}) & =\min\left\{ 1,\frac{\pi_{AB}\,Q(\mathbf{y})}{\pi_{C}}\right\} 
\end{aligned}
\label{eq:acceptance}
\end{equation}
which fulfils the given detailed balance condition~(\ref{eq:db-reduced}).
For a practical implementation one needs to know both proposal densities
$q^{+}(\mathbf{y}|\mathbf{x})$ and $q^{-}(\mathbf{x}|\mathbf{y})$,
and the corresponding acceptance probabilities $\alpha^{+}(\mathbf{y}|\mathbf{x})$
and $\alpha^{-}(\mathbf{x}|\mathbf{y})$.

\subsection{Apply DB to Doi model}

\label{sec:specific-model} 

Assuming the Doi model~(\ref{eq:lambda-on-doi},~\ref{eq:lambda-off-doi}),
we state the association proposal density $q^{+}(\mathbf{y}|\mathbf{x})$
and derive the dissociation proposal density $q^{-}(\mathbf{x}|\mathbf{y})$~(\ref{eq:proposal-reverse}).
The normalized association proposal density reads 
\begin{equation}
q^{+}(\mathbf{y}|\mathbf{x})=V^{-2}\delta\left(\mathbf{y}_{c}-\frac{\mathbf{x}_{a}+\mathbf{x}_{b}}{2}\right)\label{eq:fusion-proposal-density-doi}
\end{equation}
where the Dirac delta function $\delta(\cdot)$ assures that the $\mathrm{C}$
particle'sproposed position $\mathbf{y}_{c}$ is in the middle between
the $\mathrm{A}$ and $\mathrm{B}$ particles from the initial positions
$\mathbf{x}$. The volume term $V^{-2}$ is required for normalization,
due to the measure $\mathrm{d}\mathbf{y}=\mathrm{d}\mathbf{y}_{a}\mathrm{d}\mathbf{y}_{b}\mathrm{d}\mathbf{y}_{c}$.
Additionally the volume term can be understood as a uniform placement
of $\mathrm{A}$ and \textbf{$\mathrm{B}$ }in the final positions
$\mathbf{y}$. Since $\mathrm{A}$ and $\mathrm{B}$ are not considered
in the associated state, it is irrelevant where they are. Hence Eq.~(\ref{eq:fusion-proposal-density-doi})
fulfils $\int q^{+}(\mathbf{y}|\mathbf{x})\mathrm{d}\mathbf{y}=1$.
The normalization $Q$ of the dissociation proposal density from Eq.~(\ref{eq:q-norm-general})
can be evaluated and reduces to a constant (see Appendix C) 
\begin{equation}
Q=\frac{\lambda_{\mathrm{on}}}{k_{\mathrm{off}}}\frac{V_{\mathrm{reac}}^{\mathrm{eff}}}{V-V_{\mathrm{ex}}}.\label{eq:q-doi-model}
\end{equation}
The dissociation proposal density (\ref{eq:proposal-reverse}) then
becomes 
\begin{equation}
\begin{aligned}q^{-}(\mathbf{x}|\mathbf{y})= & \left(VV_{\mathrm{reac}}^{\mathrm{eff}}\right)^{-1}\delta\left(\mathbf{y}_{c}-\frac{\mathbf{x}_{a}+\mathbf{x}_{b}}{2}\right)\ldots\\
 & \times\chi_{\mathrm{reac}}(r)e^{-\beta U(r)},
\end{aligned}
\label{eq:dissociation-proposal-density-doi}
\end{equation}
with $r=|\mathbf{x}_{b}-\mathbf{x}_{a}|$. This density can be read
as: given a $\mathrm{C}$ particle at position $\mathbf{y}_{c}$,
positions $\mathbf{x}_{a}$ and $\mathbf{x}_{b}$ of particles $\mathrm{A}$
and $\mathrm{B}$ are restricted to radial shells concentric around
$\mathbf{y}_{c}$ due to the delta function. These shells must not
be larger than the reaction radius due to the indicator function.
The distance is additionally weighted with the Boltzmann factor of
the interaction potential $U$.

Using the normalization constant $Q$ from Eq.~(\ref{eq:q-doi-model})
the acceptance probabilities from Eq.~(\ref{eq:acceptance}) are
directly obtained. Using the microscopic association rate given in
Eq.~(\ref{eq:choice-on-rate}) results in an acceptance probability
of unity in both directions 
\[
\alpha^{+}(\mathbf{y}|\mathbf{x})=\alpha^{-}(\mathbf{x}|\mathbf{y})=1\quad\text{for~}\lambda_{\mathrm{on}}=\tilde{\lambda}_{\mathrm{on}}\text{~from\,(\ref{eq:choice-on-rate})}.
\]

\subsection{Generalize for other types of reactions\label{subsec:other-reactions}}

The presented Metropolis-Hastings Monte Carlo method can be performed
for other types of reversible reactions, namely reversible conversion
reactions of the type
\begin{equation}
\underbrace{\mathrm{A}}_{\mathbf{x}}\underset{k_{\mathrm{off}}}{\overset{k_{\mathrm{on}}}{\rightleftarrows}}\underbrace{\mathrm{B}}_{\mathbf{y}}\quad\text{with }\lambda^{+}(\mathbf{x}),\lambda^{-}(\mathbf{y})\label{eq:conversion}
\end{equation}
as well as reversible enzymatic reactions of the type
\begin{equation}
\underbrace{\mathrm{A}+\mathrm{C}}_{\mathbf{x}}\underset{k_{\mathrm{off}}}{\overset{k_{\mathrm{on}}}{\rightleftarrows}}\underbrace{\mathrm{B}+\mathrm{C}}_{\mathbf{y}}\quad\text{with }\lambda^{+}(\mathbf{x}),\lambda^{-}(\mathbf{y})\text{ and }R_{\mathrm{reac}},\label{eq:enzymatic}
\end{equation}
with macroscopic forward and backward rates $k_{\mathrm{on}}$ and
$k_{\mathrm{off}}.$ For those two reactions we can also construct
a microscopic probability density for positions $\mathbf{x}$ and
$\mathbf{y}$ for the dilute case in the fashion of Eq.~(\ref{eq:micro-distr}).
Here the microscopic phase space only has positions for $\mathrm{A}$
and $\mathrm{B}$ particles, the $\mathrm{C}$ particle in reaction~(\ref{eq:enzymatic})
can be placed at the origin without loss of generality. The reaction
functions $\lambda^{+}(\mathbf{x})$ and $\lambda^{-}(\mathbf{y})$
for the conversion reaction~(\ref{eq:conversion}) are constants
$\lambda_{\mathrm{on}}$ and $\lambda_{\mathrm{off}}$ respectively.
For the enzymatic reaction~(\ref{eq:enzymatic}) both reaction functions
are additionally multiplied with an indicator function depending on
the reaction radius $R_{\mathrm{reac}}$. As in Sec.~\ref{subsec:computing-microscopic}
we can compute the microscopic rate constants $\lambda$ that reproduce
the macroscopic kinetics in the dilute limit. In the case of the enzymatic
reaction~(\ref{eq:enzymatic}), there appear excluded volumes $V_{\mathrm{ex,A}}$,
$V_{\mathrm{ex,B}}$ and effective reaction volumes $V_{\mathrm{reac,A}}^{\mathrm{eff}}$,
$V_{\mathrm{reac,B}}^{\mathrm{eff}}$. These are defined analogously
to the volumes $V_{\mathrm{ex}}$, see Eq.~(\ref{eq:v-ex}), and
$V_{\mathrm{reac}}^{\mathrm{eff}}$, see Eq.~(\ref{eq:v-reac-eff-doi}),
with the difference that $V_{\mathrm{ex,A}}$ and $V_{\mathrm{reac,A}}^{\mathrm{eff}}$
are calculated based on the interaction potential of $\mathrm{A}$
and $\mathrm{C}$, and $V_{\mathrm{ex,B}}$ and $V_{\mathrm{reac,B}}^{\mathrm{eff}}$
are calculated based on the interaction potential of $\mathrm{B}$
and $\mathrm{C}$. To assure detailed balance we make the same ansatz
for transition rates as in Eq.~(\ref{eq:split}). The proposal densities
$q$ are constructed much simpler, because in these types of reactions
no new positions must be generated, i.e. the $q$ are delta functions.
However during the species conversion, molecules might be subject
to potentials with respect to educt and product states. We gather
the change of potential energy during the reaction in the variable
$\Delta E$. We summarize all of these findings in Tab.~\ref{tab:summary}.

\section{Results}

\label{sec:results}We have proposed a method of executing reversible
reactions according to detailed balance. It can be used to perform
reactions in a stochastic reaction-diffusion simulation. A schematic
implementation is shown in the pseudo code Alg.~\ref{alg:reaction-diffusion}.

In order to illustrate our method, we perform many-particle simulations
with molecular species $\mathrm{A}$, $\mathrm{B}$ and $\mathrm{C}$
engaging in the reversible association reaction shown in Eq.~(\ref{eq:bimol-reaction}).
The simulation is performed using overdamped Langevin dynamics in
the particle interaction potential with a fixed time-step integrator.
The potential $U(r)$ between the particles $\mathrm{A}$ and $\mathrm{B}$
is chosen as a harmonic repulsion with cutoff $R_{\mathrm{int}}$
and force constant $\kappa$, that only depends on the distance $r=|\mathbf{x}_{b}-\mathbf{x}_{a}|$
between $\mathrm{A}$ and $\mathrm{B}$ 
\begin{equation}
U(r)=\left\{ \begin{array}{rl}
\frac{1}{2}\kappa(r-R_{\mathrm{int}})^{2}, & \mathrm{if}~r<R_{\mathrm{int}}\\
0, & \mathrm{otherwise}
\end{array}\right.\label{eq:harmonic-repulsion}
\end{equation}
For this choice of potential the effective interaction volume from
Eq.~(\ref{eq:v-int-eff}) yields an expression containing errorfunctions.
In general the effective interaction volume can be determined numerically.

During one time step of length $\tau$, we first integrate the diffusive
motion of particles and then perform the reactions. The boundaries
of the system are periodic, obeying the minimum image convention and
wrapping positions upon crossing the border.

In the reaction step all possible reaction events are determined,
this depends on the considered reactions, reaction radii and the current
particle configuration. Then the list of reaction events is processed.
An event is selected from the list. The event will be proposed with
absolute probability $p=1-\exp(-\lambda\tau)$ depending on the microscopic
rate constant $\lambda$ of the associated type of reaction. The event
is performed, generating another particle configuration drawn from
the proposal densities in Eqs.~(\ref{eq:fusion-proposal-density-doi},~\ref{eq:dissociation-proposal-density-doi}).
From the change in potential energy and the type of reaction the acceptance
probability $a$ is calculated. If the event is accepted the new configuration
is kept. If the event is rejected the old configuration has to be
restored. Then the processed event is removed from the list of events.
Additionally any event is removed that would propose an event with
the same particles as the processed one, since these might not exist
anymore.

The total probability of performing a particular event is $pa$. If
$a$ is chosen according to Eq.~(\ref{eq:acceptance}) and Eq.~(\ref{eq:q-doi-model})
and the proposal density of the dissociation reaction includes the
Boltzmann factor as in Eq.~(\ref{eq:dissociation-proposal-density-doi}),
we will refer to this as the proposed \textbf{DB} reaction scheme.
We refer to the \textbf{Doi} reaction scheme if $a=1$, regardless
of the energy difference, and if the proposal density does not include
the Boltzmann factor of the interaction potential of the reactants.

\subsection{Dilute limit}

We validate Alg.~\ref{alg:reaction-diffusion} by performing it on
the system of particles $\mathrm{A}$, $\mathrm{B}$ and $\mathrm{C}$.
These particles are subject to the reaction~(\ref{eq:bimol-reaction})
and a harmonic repulsion potential as in Eq.~(\ref{eq:harmonic-repulsion}).
At any point in time there is either the $\mathrm{C}$ particle or
two particles $\mathrm{A}$ and $\mathrm{B}$, i.e. there is only
one instance of each molecule species. Thus these simulations are
in the dilute limit. The only interactions occur between the $\mathrm{A}$
and $\mathrm{B}$ particle.

\subsubsection{Validation of reaction kinetics}

\begin{figure}
\centering \includegraphics{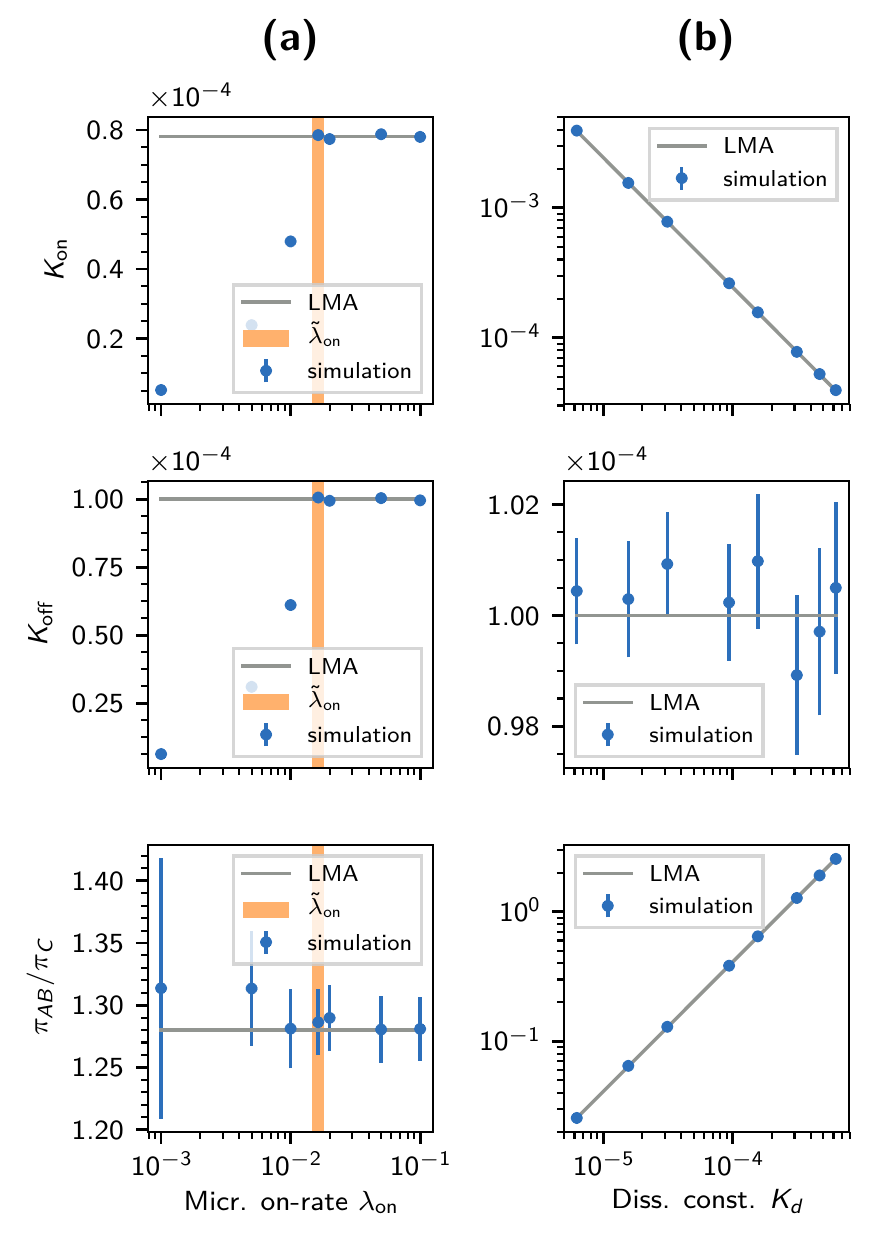} \caption{Validation of the proposed detailed balance reaction scheme in dilute
systems by stochastic particle-based reaction-diffusion simulations
(see Alg.~\ref{alg:reaction-diffusion}). Shown are observables of
the macroscopic reaction kinetics: the effective association rate
$K_{\mathrm{on}}$, the effective dissociation rate $K_{\mathrm{off}}$
and the equilibrium constant $\pi_{AB}/\pi_{C}$. Reference values
(law of mass action - LMA) for $K_{\mathrm{on}}$, $K_{\mathrm{off}}$
and $\pi_{AB}/\pi_{C}$ correspond to macroscopic behaviour described
in Sec.~\ref{sec:bi-mol-equilibrium}. See simulation parameters
in Tab.~\ref{tab:parameters-dilute}. \textbf{(a)}~Microscopic association
rate constant $\lambda_{\mathrm{on}}$ is varied. $\tilde{\lambda}_{\mathrm{on}}$
corresponds to Eq.~(\ref{eq:choice-on-rate}). \textbf{(b)}~The
given dissociation constant $K_{d}$ is varied. The microscopic association
rate constant is $\lambda_{\mathrm{on}}=\tilde{\lambda}_{\mathrm{on}}(K_{d})$. }
\label{fig:dilute-macro-kinetics} 
\end{figure}

We show that the proposed detailed balance reaction scheme always
yields the desired macroscopic equilibrium distribution $\pi_{AB}/\pi_{C}$
from Eq.~(\ref{eq:equilibrium-constant}). Additionally we demonstrate
under which circumstances the simulated effective on- and off-rates,
$K_{\mathrm{on}}$ and $K_{\mathrm{off}}$, will match those given
by Eq.~(\ref{eq:macroscopic-effective-on-rate}) and Eq.~(\ref{eq:macroscopic-effective-off-rate}).
The results are seen in Fig.~\ref{fig:dilute-macro-kinetics}. The
simulation parameters are given in Tab.~\ref{tab:parameters-dilute}.

Fig.~\ref{fig:dilute-macro-kinetics}a shows that for very low $\lambda_{\mathrm{on}}$,
the effective association rate $K_{\mathrm{on}}$ cannot exceed a
certain value because the proposal frequency is limited and $K_{\mathrm{off}}$
is in turn diminished by rejection of dissociation events in order
to reproduce the desired equilibrium constant $\pi_{AB}/\pi_{C}=K_{d}V$.
For very high $\lambda_{\mathrm{on}}$, association events will be
rejected, thus limiting $K_{\mathrm{on}}$ to the LMA value, while
dissociation events are executed with frequency $K_{\mathrm{off}}=k_{\mathrm{off}}$.
The transition between these two regimes is where $\lambda_{\mathrm{on}}=\tilde{\lambda}_{\mathrm{on}}$
as in Eq.~(\ref{eq:choice-on-rate}). Fig.~\ref{fig:dilute-macro-kinetics}b
shows that, when one uses the appropriate association rate constant
from Eq.~(\ref{eq:choice-on-rate}), one can reproduce the expected
reaction kinetics for varying $K_{d}$.

\subsubsection{Microscopic reversibility}

\label{sec:microscopic-reversibility} 
\begin{figure}[htb]
\centering \includegraphics{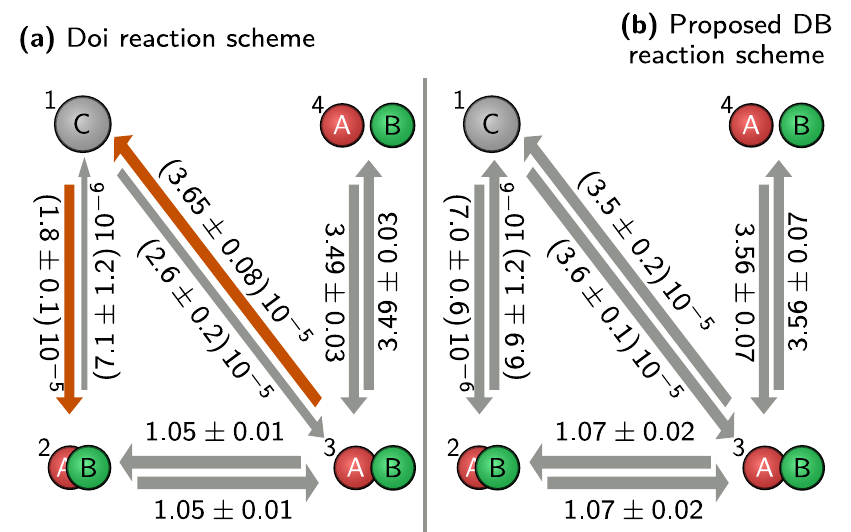}
\caption{Probability fluxes between associated and dissociated states measured
from particle-based reaction-diffusion simulations (see Alg.~\ref{alg:reaction-diffusion})
in the dilute limit. Compared are the Doi reaction scheme and the
proposed detailed balance reaction scheme (DB). Definitions of the
states 1-4 are given in Sec.~\ref{sec:microscopic-reversibility}.
Arrows depict transitions between these states as observed in the
simulations. The width of the arrows encodes the probability flux
$\pi_{i}K_{ij}$, also given as numeric values measured from multiple
independent simulations giving rise to the standard error of the mean.
The widths of two adjacent arrows are normalized with respect to each
other (not globally). See parameters in Tab.~\ref{tab:parameters-dilute}.
\textbf{(a)}~Doi reaction scheme. The probability fluxes for the
transitions $1\to2$ and $1\to3$ are imbalanced compared to their
respective counterparts, resulting in a circular flux of probability.
\textbf{(b)}~Detailed balance reaction scheme. }
\label{fig:microscopic-reversibility} 
\end{figure}

We now demonstrate that the
proposed DB reaction scheme (Alg.~\ref{alg:reaction-diffusion})
indeed produces trajectories in thermodynamic equilibrium, while the
naive Doi scheme leads to periodic cycles in phase space, corresponding
to an unintended nonequilibrium scenario. To this end, we distinguish
three substates of the dissociated state, defined by the inter-particle
distance $r$ of particles $\mathrm{A}$ and $\mathrm{B}$, and the
reaction radius $R$. We define states 1-4 as follows:
\begin{enumerate}
\item The complex state, $\mathrm{C}$ 
\item $\mathrm{A}$ and $\mathrm{B}$ are very close $r\leq\frac{3}{4}R$ 
\item $\mathrm{A}$ and $\mathrm{B}$ are still in reactive range $\frac{3}{4}R<r\leq R$ 
\item $\mathrm{A}$ and $\mathrm{B}$ are not within reactive range $r>R$ 
\end{enumerate}
Using again a reversibly reacting system with a single $\mathrm{A},\mathrm{B}$
pair or a single $\mathrm{C}$ complex, we determine the stationary
distribution $\pi$ for this definition of states, and the transition
rates $K$ connecting them. A process that fulfils detailed balance
must yield 
\begin{equation}
\pi_{i}K_{ij}=\pi_{j}K_{ji}\label{eq:microscopic-reversibility}
\end{equation}
for all pairs of states $i,j$. We measure $\pi$ and $K$ from simulations
and compare the Doi reaction scheme and the proposed DB reaction scheme
in the presence of a harmonic repulsion potential between $\mathrm{A}$
and $\mathrm{B}$. In this comparison all system parameters are identical,
only the reaction mechanism differs. Results are presented in Fig.~\ref{fig:microscopic-reversibility}
and simulation parameters are given in Tab.~\ref{tab:parameters-dilute}.

From Fig.~\ref{fig:microscopic-reversibility} it is evident that
for the present case of interacting particles, the naive Doi reaction
scheme produces a cyclic probability flux that violates DB. In the
proposed DB reaction scheme, this is not the case and all given probability
fluxes obey Eq.~(\ref{eq:microscopic-reversibility}).

Note that for both reaction schemes, there occurs a unidirectional
transition $4\to1$ due to the time splitting we employ during one
simulation step (first the diffusion step and then the reaction step).
This artificial transition is a result of the time-step discretization
error and not related to the DB scheme. It occurs with an absolute
rate of less than $10^{-6}$, all other transitions have $K_{ij}>10^{-5}\forall(i,j)\neq(4,1)$.
Thus its probability flux is not shown here.

\subsection{System of many particles}

Finally, we study how a dense mixture of interacting particles behaves
when the DB algorithm is employed, and we compare this behavior with
the naive Doi algorithm and what is expected from physical intuition.
The Algorithm~\ref{alg:reaction-diffusion} is performed for a system
of many $\mathrm{A}$, $\mathrm{B}$ and $\mathrm{C}$ particles confined
to the volume $V$ with periodic boundaries. In this scenario we assign
physical radii $r_{A}$, $r_{B}$, and $r_{C}$ to the particles.
Particles are subject to harmonic repulsion potentials~(\ref{eq:harmonic-repulsion})
acting between all pairs of species $\mathrm{A}$, $\mathrm{B}$,
and $\mathrm{C}$, where the interaction radius is chosen as the sum
of the particles' radii. See parameters in Tab.~\ref{tab:parameters-dense}.
Particles are subject to the reaction~(\ref{eq:bimol-reaction}).
Employing the DB reaction scheme introduced in Sec.~\ref{sec:db-general}
can therefore result in rejected Monte-Carlo moves, which will affect
the thermodynamics and kinetics of the simulation system in the dense
limit.

In Sec.~\ref{sec:microscopic-distribution} and following we had
assumed that phase space consists of only three particles $\mathrm{A}$,
$\mathrm{B}$ and $\mathrm{C}$. In the case of many possible reactants
one is presented with multiple possible reaction events. For one particular
event we will use the proposal densities from Eqs.~(\ref{eq:fusion-proposal-density-doi},~\ref{eq:dissociation-proposal-density-doi})
to treat the particles taking part in the event. All other particles
will be considered static excess objects. This means that the microscopic
distributions from Eq.~(\ref{eq:micro-distr}) gain another Boltzmann
factor from interactions with the excess particles. Note that the
partition functions $Z_{AB}$ and $Z_{C}$ will differ from their
``dilute'' values. In Sec.~\ref{sec:specific-model} we have seen
that a particular choice of parameters leads to the prefactor in the
acceptance becoming unity. Hence, the advantage of such a Markov Chain
Monte Carlo algorithm is that one does not need to know constant factors
of the stationary distribution to draw samples from said distribution.
Along these lines we construct an acceptance function for the many
particle case, that includes a Boltzmann factor of the energy difference
and a prefactor of unity, assuming that internal reaction parameters
correspond to a certain but unknown macroscopic equilibrium. We will
use the association rate constant derived in Eq.~(\ref{eq:choice-on-rate}).
Obviously this equilibrium will differ from the one in Eq.~(\ref{eq:equilibrium-constant}).
But one can guarantee detailed balance never the less.

The change of potential energy is $\Delta\epsilon$. It does not include
the interaction between $\mathrm{A}$ and $\mathrm{B}$ as this is
already accounted for by the proposal probabilities $q^{+}$ and $q^{-}$.
We may write $\Delta\epsilon$ as the total change of potential energy
$\Delta E$ minus the interaction energy $U_{AB}$. We formulate the
acceptance for the many particle case:
\begin{equation}
\begin{aligned}\alpha^{+}(\mathbf{y}|\mathbf{x}) & =\min\left\{ 1,\exp(-\beta\Delta\epsilon^{+})\right\} \\
\alpha^{-}(\mathbf{x}|\mathbf{y}) & =\min\left\{ 1,\exp(-\beta\Delta\epsilon^{-})\right\} 
\end{aligned}
\label{eq:acceptance-many}
\end{equation}
where the changes of energies are given by 
\begin{equation}
\begin{aligned}\Delta\epsilon^{+} & =E(\mathbf{y})-[E(\mathbf{x})-U_{AB}(\mathbf{x})]\\
\Delta\epsilon^{-} & =[E(\mathbf{x})-U_{AB}(\mathbf{x})]-E(\mathbf{y}).
\end{aligned}
\label{eq:energy-difference-many}
\end{equation}
We set up the system with a certain number of $\mathrm{A}$ and $\mathrm{B}$
particles and no $\mathrm{C}$ particles. We control the quantity
$n=(N_{A}+N_{B})/2+N_{C}$ which is conserved during a simulation.
The system equilibrates without the reaction, we then switch the reaction
on and let the system equilibrate again.

We compute three observables in the equilibrated state, i.e. when
observables are stable and converged: the equilibrium constant $\pi_{AB}/\pi_{C}=V[A][B]/[C]$,
the total potential energy of the system $U$ in units of $k_{B}T$
and the pressure $P$ in units of $V^{-1}k_{B}T$. The pressure is
measured from evaluating the virial term of acting forces as described
in~\citep{Allen1987}. Individual reactions are integrated with either
the proposed DB scheme or the Doi reaction scheme.

\begin{figure}
\centering\includegraphics[width=3.5in]{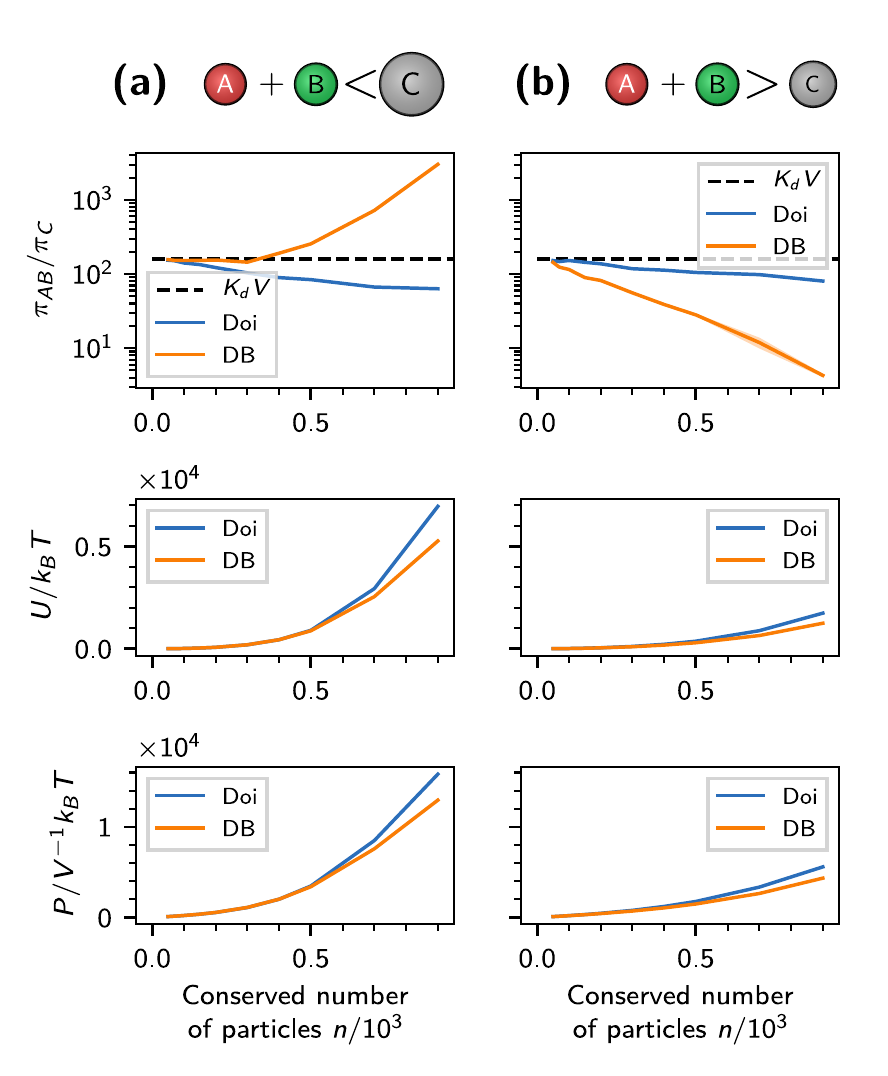}\caption{Steady state observables measured in particle-based reaction-diffusion
simulations with multiple particles. The quantity $n=(N_{A}+N_{B})/2+N_{C}$
is conserved during a simulation. Shown are ensemble- and time-averaged
values of the equilibrium constant $\pi_{AB}/\pi_{C}=V[A][B]/[C]$,
the potential energy $U$ in units of $k_{B}T$, the pressure $P$
in units of $V^{-1}k_{B}T$. Compared are the two reaction schemes
Doi and DB, see Sec.~\ref{sec:results}. See simulation parameters
in Tab.~\ref{tab:parameters-dense} \textbf{(a)}~An association
reaction of $\mathrm{A}$ and $\mathrm{B}$ increases the total volume
occupied by particles such that $r_{A}^{3}+r_{B}^{3}<r_{C}^{3}$.
\textbf{(b)}~The $\mathrm{C}$ particle occupies less volume than
A and B combined such that $r_{A}^{3}+r_{B}^{3}>r_{C}^{3}$ }

\label{fig:dense-observables} 
\end{figure}

\begin{figure*}[t]
\centering \includegraphics[width=1\textwidth]{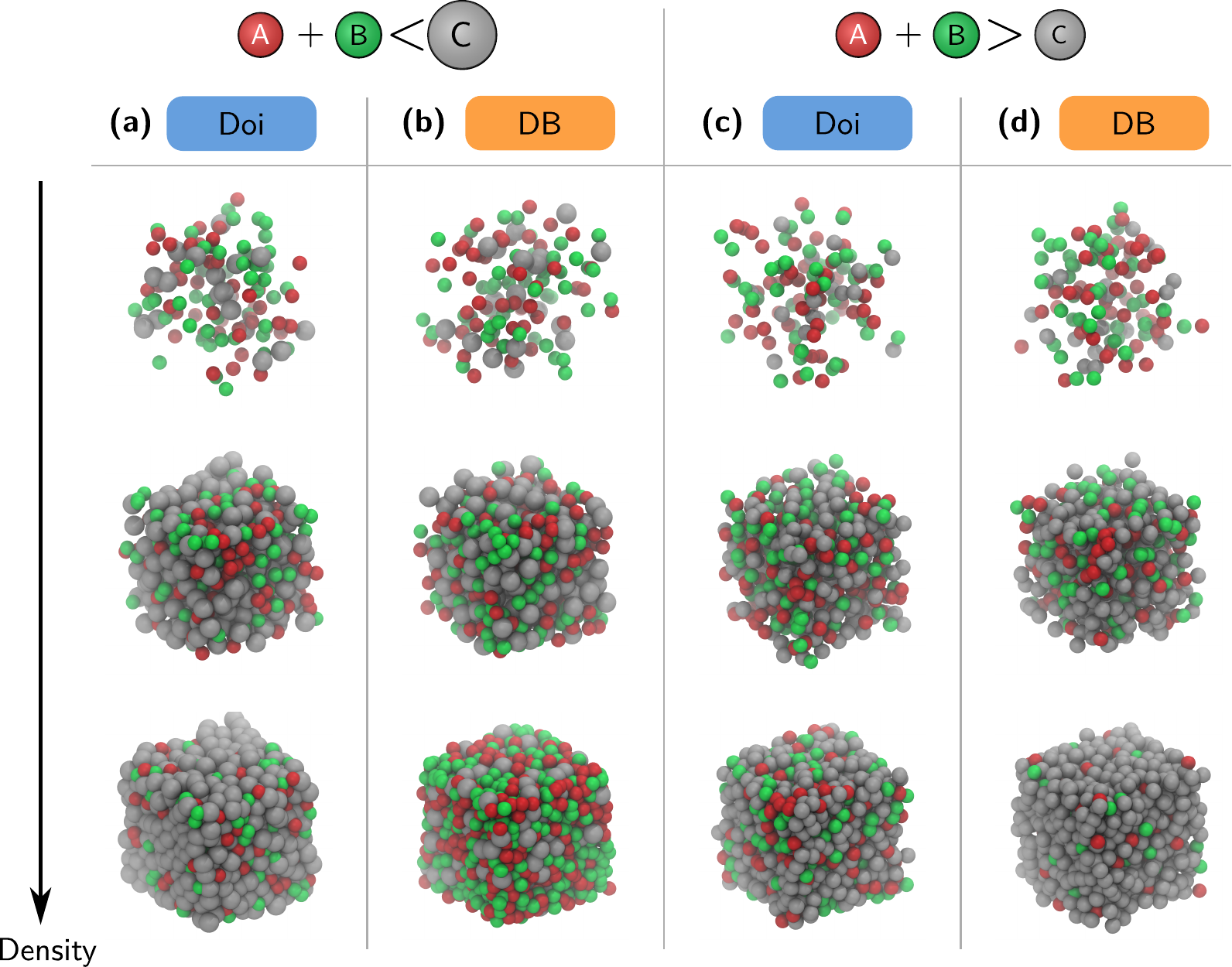}
\caption{Steady state configurations of particle-based reaction-diffusion simulations
subject to the reaction $\mathrm{A}+\mathrm{B}\rightleftarrows\mathrm{C}$
for different densities in terms of the number of particles $n$ initially
in the system. Compared are the two reaction schemes Doi and DB, see
Sec.~\ref{sec:results} at different particle radii respectively.
See simulation parameters in Tab.~\ref{tab:parameters-dense}. \textbf{(a)~}The
associated state occupies more volume than the dissociated state,
reactions are handled with the Doi scheme. \textbf{(b)}~The associated
state occupies more volume than the dissociated state, reactions are
handled with the DB scheme \textbf{(c)~}The associated state occupies
less volume than the dissociated state, reactions are handled with
the Doi scheme. \textbf{(d)}~The associated state occupies less volume
than the dissociated state, reactions are handled with the DB scheme.}
\label{fig:dense-snapshots} 
\end{figure*}

Fig.~\ref{fig:dense-observables}a shows the results for the case
when an association reaction of $\mathrm{A}$ and $\mathrm{B}$ increases
the total volume occupied by particles such that $r_{A}^{3}+r_{B}^{3}<r_{C}^{3}$.
The associated state is energetically less favourable. In the dilute
limit both methods Doi and DB reproduce the macroscopic equilibrium
population $\pi_{AB}/\pi_{C}=K_{d}V$. For increasing number of particles
both methods differ significantly. The Doi reaction scheme favours
the energetically higher associated configuration $\mathrm{C}$. The
Doi scheme produces an equilibrium constant of roughly $\pi_{AB}/\pi_{C}\approx80$
for the highest density simulated. The DB scheme adjusts the effective
association probability by rejecting association events. This results
in a steady state, where almost no $\mathrm{C}$ particles exist with
an equilibrium constant exceeding $\pi_{AB}/\pi_{C}>3\times10^{3}$.
For all $n>50$, the DB scheme finds a steady state of lower energy
and lower pressure compared to the Doi scheme. Fig.~\ref{fig:dense-snapshots}a
and b show representative simulation snapshots of the steady states
for Doi and DB scheme.

Fig.~\ref{fig:dense-observables}b shows the case when a $\mathrm{C}$
particle occupies less volume than $\mathrm{A}$ and $\mathrm{B}$
combined such that $r_{A}^{3}+r_{B}^{3}>r_{C}^{3}$, which could correspond
to two proteins $\mathrm{A}$ and $\mathrm{B}$, which only fully
fold in a bound state. In the dilute case both methods Doi and DB
reproduce the same behaviour in all three observables. For increasing
number of particles the Doi method produces a similar steady state
population as in Fig.~\ref{fig:dense-observables}a where the $C$
state is favoured. The DB scheme produces states favouring the $C$
state even stronger thus reducing the system's potential energy and
pressure compared to the Doi scheme. Fig.~\ref{fig:dense-snapshots}c
and d show representative simulation snapshots of the steady states
for Doi and DB scheme.

\section{Conclusion}

We have derived an algorithm to perform iPRD simulations of molecules
undergoing reversible reactions of the form $\mathrm{A}+\mathrm{B}\rightleftarrows\mathrm{C}$
according to detailed balance. This method is called iPRD-DB. 

Detailed balance guarantees that simulations of an isolated system
generate samples according to thermodynamic equilibrium. We have shown
that in a dense reactive mixture of particles, that exhibit volume
exclusion due to pair-wise potentials, the steady state of the system
simulated with iPRD-DB is in agreement with Henri Le Chatelier's principle~\citep{Atkins2006},
i.e. that the achieved steady state concentrations strongly depend
on the interaction of molecules. Biochemical pathways often show switch-like
behavior, and are thus sensitive to such changes in concentrations
of agents~\citep{Nuoffer1994,hall2000rho,Marat2017}. Sampling the
correct equilibrium is crucial when simulating such processes.

The iPRD-DB method can be generalized for other types of reactions,
such as a reversible change of molecule species $\mathrm{A}\rightleftarrows\mathrm{B}$,
or a reversible enzymatic reaction $\mathrm{A}+\mathrm{C}\rightleftarrows\mathrm{B}+\mathrm{C}$,
which describes a Michaelis-Menten experiment when the backwards rate
becomes very small.

Furthermore the iPRD-DB method is accompanied by an equation for the
microsopic rate constant $\lambda$ that assures the correct macroscopic
reaction kinetics. This equation, see Eq.~(\ref{eq:choice-on-rate}),
relates the macroscopic kinetic parameters $K_{d}$ and $k_{\mathrm{off}}$
in a dilute environment with the microscopic iPRD model parameters:
microscopic rate constant $\lambda$, reaction radius $R$, and force
parameters that determine the excluded volume $V_{\mathrm{ex}}$.
Thus, it provides a choice for $\lambda$, which in the iPRD-DB algorithm
functions as the absolute proposal rate. For this choice the acceptance
probability reduces to the Boltzmann factor describing the change
of energy with respect to educt and product states. We also provide
proposal densities such that the acceptance becomes unity in the dilute
case.

Having measured $K_{d}$ and $k_{\mathrm{off}}$ in an \textit{in
vitro} scenario, a microscopic iPRD model can be constructed subject
to Eq.~(\ref{eq:choice-on-rate}) and can then be analyzed numerically
to gain insights about the \textit{in vivo} process, where molecules
may occur in very low copy numbers and diffuse anomalously due to
complex geometries, making experimental measurements cumbersome in
this regime. Note that the expression relating $K_{d}$ and $k_{\mathrm{off}}$
with $\lambda$ and $R$ is independent of the diffusion coefficient
$D$, i.e. an iPRD model can be adjusted to resemble the in vivo effective
diffusion, which may, e.g. be obtained from Förster resonance energy
transfer (FRET) experiments~\citep{Lippincott-Schwartz2003}.

An open question is what the analytical reference chemical equilibrium
is when going to dense particle mixtures.

\section*{Acknowledgements}

This paper is dedicated to William A. Eaton on the occasion of his
80th birthday. Happy Birthday Bill - you are a wonderful person and
your work has been an inspiration for this field!

We gratefully acknowledge funding from Deutsche Forschungsgemeinschaft
(SFB 958 / Project A04, TRR 186 / Project A12, SFB 1114 / Project
C03), Einstein Foundation Berlin (ECMath Project CH17) and European
Research Council (ERC CoG 772230 ``ScaleCell''). We are grateful
for inspiring discussions with Moritz Hoffmann, Manuel Dibak, Luigi
Sbailò, Mohsen Sadeghi, Felix Höfling and Christof Schütte.

\section*{Appendix}

\subsection*{A. Normalization constant $Z_{AB}$}

The normalization is 
\[
\begin{aligned}Z_{AB}= & \int e^{-\beta U(\mathbf{x})}\mathrm{d}\mathbf{x}\\
= & \int\mathrm{d}\mathbf{x}_{c}\iint e^{-\beta U(\mathbf{x}_{b}-\mathbf{x}_{a})}\mathrm{d}\mathbf{x}_{a}\mathrm{d}\mathbf{x}_{b}\\
= & V\left(I_{1}+I_{2}\right)
\end{aligned}
\]
If there are no external potentials present, the latter integral factorizes
\[
\begin{aligned}I_{2} & =\iint_{|\mathbf{x}_{b}-\mathbf{x}_{a}|>R_{\mathrm{int}}}\mathrm{d}\mathbf{x}_{a}\mathrm{d}\mathbf{x}_{b}\\
 & =\int\left(\int_{|\mathbf{x}_{b}-\mathbf{x}_{a}|>R_{\mathrm{int}}}\mathrm{d}\mathbf{x}_{b}\right)\mathrm{d}\mathbf{x}_{a}\\
 & =\left(V-V_{\mathrm{int}}\right)\int\mathrm{d}\mathbf{x}_{a}=\left(V-V_{\mathrm{int}}\right)V
\end{aligned}
\]
where $V_{\mathrm{int}}$ is the interaction volume, that only depends
on the cut-off distance of the potential $R_{\mathrm{in}}$, not the
potential itself. Since the potential $U$ only depends on the relative
position $\mathbf{x}_{b}-\mathbf{x}_{a}$, one can fix the position
of one particle without changing the value of the integral $I_{1}$
\[
\begin{aligned}I_{1} & =\iint_{|\mathbf{x}_{b}-\mathbf{x}_{a}|\leq R_{\mathrm{int}}}e^{-\beta U(\mathbf{x}_{b}-\mathbf{x}_{a})}\mathrm{d}\mathbf{x}_{a}\mathrm{d}\mathbf{x}_{b}\\
 & =\int\left(\int_{|\mathbf{x}_{b}-\mathbf{x}_{a}|\leq R_{\mathrm{int}}}e^{-\beta U(\mathbf{x}_{b}-\mathbf{x}_{a})}\mathrm{d}\mathbf{x}_{b}\right)\mathrm{d}\mathbf{x}_{a}\\
 & =V_{\mathrm{int}}^{\mathrm{eff}}~\int\mathrm{d}\mathbf{x}_{a}=V_{\mathrm{int}}^{\mathrm{eff}}~V
\end{aligned}
\]
The effective accessible volume inside the interaction radius is given
by:
\[
V_{\mathrm{int}}^{\mathrm{eff}}=V_{\mathrm{int}}-V_{\mathrm{ex}},
\]
which defines the excluded volume $V_{\mathrm{ex}}$ due to interaction

\subsection*{B. Relation to diffusion-influenced rate constant derivations}

To understand Eq.~(\ref{eq:choice-on-rate}) we formulate the association
rate constant for our problem using Eq.~(\ref{eq:dissociation-constant})
\begin{equation}
k_{\mathrm{on}}=\tilde{\lambda}_{\mathrm{on}}V\frac{V_{\mathrm{reac}}^{\mathrm{eff}}}{V-V_{\mathrm{ex}}}.\label{eq:association-rate-constant}
\end{equation}
This rate is linearly dependent on the effective reaction volume from
Eq.~(\ref{eq:v-reac-eff-doi}), i.e. if one increases the repulsion
force between particles $\mathrm{A}$ and $\mathrm{B}$ the association
rate will decrease. One further notices that the diffusion of particles
is not considered in this equation, since we assume they are at all
times distributed according to Eq.~(\ref{eq:micro-distr}). This
is true only because of the reversible reaction that the isolated
pair is subject to. The diffusion approach of $\mathrm{A}$ and $\mathrm{B}$
need not be considered here. It is therefore crucial in an algorithm
to generate samples from the stationary distribution we assumed.

At this point we can establish a connection with other treatments
of diffusion influenced reaction rates. The formula derived by Doi~\citep{Doi1975}
describes the association rate constant for particles approaching
each other via diffusion from the far-field. It includes the relative
diffusion constant of the two particles $D$ and reads
\[
k_{\mathrm{on,Doi}}=4\pi DR\left(1-\sqrt{\frac{D}{\lambda_{\mathrm{on}}R^{2}}}\tanh\left(\sqrt{\frac{\lambda_{\mathrm{on}}R^{2}}{D}}\right)\right)
\]
Assuming the fast diffusion limit of this yields~\citep{Erban2009}
\begin{equation}
\lambda\ll\frac{D}{R^{2}}\quad\to\quad k_{\mathrm{on,Doi}}\approx\lambda_{\mathrm{on}}\frac{4}{3}\pi R^{3}.\label{eq:erb-chap-rate-limit}
\end{equation}
If we on the other hand assume the large volume limit of the expression
from Eq.~(\ref{eq:association-rate-constant}) we arrive at 
\begin{equation}
R^{3}\ll V\quad\to\quad k_{\mathrm{on}}=\tilde{\lambda}_{\mathrm{on}}V_{\mathrm{reac}}^{\mathrm{eff}}.\label{eq:association-rate-constant-large-volume}
\end{equation}
Comparing Eqs.~(\ref{eq:erb-chap-rate-limit},\ref{eq:association-rate-constant-large-volume})
we see that they match if the term $4\pi R^{3}/3$ is identified as
the effective reaction volume without potentials. 

\subsection*{C. Normalization of dissociation proposal density}

Additionally we need $Q(\mathbf{y})$ from \ref{eq:q-norm-general}
\[
\begin{aligned}Q(\mathbf{y})= & \frac{\lambda_{\mathrm{on}}V}{k_{\mathrm{off}}Z_{AB}}\iiint\delta\left(\mathbf{y}_{c}-\frac{\mathbf{x}_{a}+\mathbf{x}_{b}}{2}\right)\chi_{\mathrm{reac}}(\mathbf{x})\ldots\\
 & \times e^{-\beta U(|\mathbf{x}_{b}-\mathbf{x}_{a}|)}\mathrm{d}\mathbf{x}_{a}\mathrm{d}\mathbf{x}_{b}\mathrm{d}\mathbf{x}_{c}\\
= & \frac{\lambda_{\mathrm{on}}V^{2}}{k_{\mathrm{off}}Z_{AB}}\iint\displaylimits_{|\mathbf{x}_{b}-\mathbf{x}_{a}|\leq R}\delta\left(\mathbf{y}_{c}-\frac{\mathbf{x}_{a}+\mathbf{x}_{b}}{2}\right)\ldots\\
 & \times e^{-\beta U(|\mathbf{x}_{b}-\mathbf{x}_{a}|)}\mathrm{d}\mathbf{x}_{a}\mathrm{d}\mathbf{x}_{b}
\end{aligned}
\]
The delta function can be reformulated in relative coordinates of
$\mathrm{A}$ and $\mathrm{B}$, that have to placed symmetric around
$\mathbf{y}_{c}$. This eliminates another integral, which yields
$1$, due to the delta function. The only remaining degree of freedom
is the distance of $\mathrm{A}$ and $\mathrm{B}$, which results
in an integral, that is identical to the effective reaction volume
$V_{\mathrm{reac}}^{\mathrm{eff}}$ from Eq.~(\ref{eq:v-reac-eff-doi}).

\bibliographystyle{ieeetr}
\bibliography{literature/books,literature/papers_db,literature/in_preparation}

\SetKwFunction{randomUniform}{random-uniform}
\SetFuncSty{textsc}
\SetArgSty{}
\SetProgSty{}
\SetFuncArgSty{textsc}
\SetCommentSty{}
\SetKwProg{Fn}{Function}{ is}{end}
\begin{algorithm}
    \caption{Reaction diffusion algorithm for $n$ integration steps with time step size $\tau$}\label{alg:reaction-diffusion}
    \DontPrintSemicolon
    initialize list of particles/system state $p$\;
    \Repeat{$n$ steps performed}{
        $f\gets$ calculate forces for state $p$\;
        $p\gets$ propagate diffusion subject to $f$ and $\tau$\;
        $L\gets$ list of possible reaction events in $p$\;
        \While{$L$ not empty}{
            select next event $l$ from $L$\;
            $u_1\gets\randomUniform$\;
            $\lambda\gets$ microscopic rate constant of $l$\;
            \If{$u_1 < 1-\exp(\lambda\tau)$} {
                $E_1\gets$ calculate energy of state $p$\;
                $p\gets$ propose event $l$ according to density $q$\;
                $E_2\gets$ calculate energy of state $p$\;
                $a\gets$ acceptance for $l$ and energies $E_1$ and $E_2$\;
                $u_2\gets\randomUniform$\;
                \eIf{$u_2 < a$} {
                    accepted, keep the state $p$\;
                } {
                    $p\gets$ revert the event $l$\;
                }
            }
            remove $l$ out of $L$\;
            remove all events out $L$, that shared particles with event $l$\;
        }
    }
\end{algorithm}

\begin{table}[tbh]
\centering %
\begin{tabular}{lcc}
\toprule 
Quantity  & Symbol  & Value \tabularnewline
\midrule 
Dissociation constant  & $K_{d}$  & $3.125\times10^{-4}$\tabularnewline
Dissociation rate constant  & $k_{\mathrm{off}}$  & $10^{-4}$ \tabularnewline
Volume  & $V$  & $16\times16\times16$ \tabularnewline
Diffusion constant of each particle  & $D$  & $5$\tabularnewline
Reaction radius  & $R_{\mathrm{reac}}$  & $2$ \tabularnewline
Interaction radius  & $R_{\mathrm{int}}$  & $2$ \tabularnewline
Force constant  & $\kappa$  & $5$ \tabularnewline
Time step length &  & \tabularnewline
~ in Fig.~\ref{fig:dilute-macro-kinetics}  & $\tau_{1}$  & $10^{-4}$ \tabularnewline
~ in Fig.~\ref{fig:microscopic-reversibility}  & $\tau_{2}$  & $1.25\times10^{-5}$ \tabularnewline
Number of integration steps  &  & \tabularnewline
~ in Fig.~\ref{fig:dilute-macro-kinetics}  & $m_{1}$  & $3\times10^{10}$ \tabularnewline
~ in Fig.~\ref{fig:microscopic-reversibility}  & $m_{2}$  & $4.8\times10^{11}$ \tabularnewline
\bottomrule
\end{tabular}\caption{Unitless parameters used in the simulations of dilute systems, see
Fig.~\ref{fig:dilute-macro-kinetics} and \ref{fig:microscopic-reversibility}.}
\label{tab:parameters-dilute} 
\end{table}

\begin{table}[tbh]
\centering %
\begin{tabular}{lcc}
\toprule 
Quantity  & Symbol  & Value \tabularnewline
\midrule 
Dissociation constant  & $K_{d}$  & $2\times10^{-2}$\tabularnewline
Dissociation rate constant  & $k_{\mathrm{off}}$  & $10^{-3}$ \tabularnewline
Volume & $V$  & $20\times20\times20$ \tabularnewline
Particle radii &  & \tabularnewline
~ case $r_{A}^{3}+r_{B}^{3}<r_{C}^{3}$ & $(r_{A},r_{B},r_{C})$ & $(1,1,1.4)$\tabularnewline
~ case $r_{A}^{3}+r_{B}^{3}>r_{C}^{3}$ & $(r_{A},r_{B},r_{C})$ & $(1,1,1.1)$\tabularnewline
Diffusion constants per radius &  &  \tabularnewline
~for species $i\in\{A,B,C\}$ & $D/r_{i}$  & $5$\tabularnewline
Interaction radius for pair  &  & \tabularnewline
~ of species $(i,j)\forall i,j\in\{A,B,C\}$ & $R_{\mathrm{int}}(i,j)$  & $r_{i}+r_{j}$\tabularnewline
Reaction radius  & $R_{\mathrm{reac}}$  & $2$ \tabularnewline
Force constant & $\kappa$  & $10$ \tabularnewline
Time step length & $\tau$ & $5\times10^{-4}$\tabularnewline
Time steps until equilibrated &  & \tabularnewline
~ dilute system with $n=50$ & $m_{\mathrm{dilute}}$ & $1.2\times10^{8}$\tabularnewline
~ dense system with $n=900$ & $m_{\mathrm{dense}}$ & $9\times10^{6}$\tabularnewline
\bottomrule
\end{tabular}\caption{Unitless parameters used in the simulations of dense systems, see
Fig.~\ref{fig:dense-observables} and \ref{fig:dense-snapshots}.}
\label{tab:parameters-dense} 
\end{table}

\begin{table*}
\centering%
\begin{tabular}{>{\raggedright}p{2cm}lll}
\toprule 
 & $\underbrace{\text{A}+\text{B}}\limits _{\mathbf{x}}\underset{k_{\mathrm{off}}}{\overset{k_{\mathrm{on}}}{\rightleftarrows}}\underbrace{\text{C}}\limits _{\mathbf{y}}$  & $\underbrace{A}_{\mathbf{x}}\underset{k_{\mathrm{off}}}{\overset{k_{\mathrm{on}}}{\rightleftarrows}}\underbrace{\mathrm{B}}_{\mathbf{y}}$  & $\underbrace{\mathrm{A}+\mathrm{C}}_{\mathbf{x}}\underset{k_{\mathrm{off}}}{\overset{k_{\mathrm{on}}}{\rightleftarrows}}\underbrace{\mathrm{B}+\mathrm{C}}_{\mathbf{y}}$ \tabularnewline
\midrule 
$\lambda^{+}(\mathbf{x})$  & $\lambda_{\mathrm{on}}\chi_{\mathrm{reac}}(\mathbf{x})$  & $\lambda_{\mathrm{on}}$  & $\lambda_{\mathrm{on}}\chi_{\mathrm{reac}}(\mathbf{x})$ \tabularnewline
$\lambda^{-}(\mathbf{y})$  & $\lambda_{\mathrm{off}}$  & $\lambda_{\mathrm{off}}$  & $\lambda_{\mathrm{off}}\chi_{\mathrm{reac}}(\mathbf{y})$ \tabularnewline
$q^{+}(\mathbf{y}|\mathbf{x})$  & $V^{-2}\delta\left(\mathbf{y}_{c}-\frac{\mathbf{x}_{a}+\mathbf{x}_{b}}{2}\right)$  & $\delta(\mathbf{y}-\mathbf{x})$  & $\delta(\mathbf{y}-\mathbf{x})$ \tabularnewline
$q^{-}(\mathbf{x}|\mathbf{y})$  & %
\parbox[t]{5cm}{%
$\left(VV_{\mathrm{reac}}^{\mathrm{eff}}\right)^{-1}\delta\left(\mathbf{y}_{c}-\frac{\mathbf{x}_{a}+\mathbf{x}_{b}}{2}\right)$\linebreak{}
 $\ldots\times\chi_{\mathrm{reac}}(\mathbf{x})e^{-\beta U_{AB}(\mathbf{x})}$%
}  & $\delta(\mathbf{x}-\mathbf{y})$  & $\delta(\mathbf{x}-\mathbf{y})$ \tabularnewline
$f^{+}(\mathbf{y}|\mathbf{x})$  & $e^{-\beta(E(\mathbf{y})-[E(\mathbf{x})-U_{AB}(\mathbf{x})])}$  & $e^{-\beta(E(\mathbf{y})-E(\mathbf{x}))}$  & $\frac{V_{\mathrm{reac,A}}^{\mathrm{eff}}}{V_{\mathrm{reac,B}}^{\mathrm{eff}}}e^{-\beta(E(\mathbf{y})-E(\mathbf{x}))}$ \tabularnewline
$f^{-}(\mathbf{x}|\mathbf{y})$  & $e^{-\beta([E(\mathbf{x})-U_{AB}(\mathbf{x})]-E(\mathbf{y}))}$  & $e^{-\beta(E(\mathbf{x})-E(\mathbf{y}))}$  & $\frac{V_{\mathrm{reac,B}}^{\mathrm{eff}}}{V_{\mathrm{reac,A}}^{\mathrm{eff}}}e^{-\beta(E(\mathbf{x})-E(\mathbf{y}))}$ \tabularnewline
\midrule
constraints & %
\parbox[t]{5cm}{%
\raggedright$k_{\mathrm{on}}=\lambda_{\mathrm{on}}V\frac{V_{\mathrm{reac}}^{\mathrm{eff}}}{V-V_{\mathrm{ex}}}$\linebreak{}
 $k_{\mathrm{off}}=\lambda_{\mathrm{off}}$ %
}  & %
\parbox[t]{3cm}{%
\raggedright$k_{\mathrm{on}}=\lambda_{\mathrm{on}}$\linebreak{}
 $k_{\mathrm{off}}=\lambda_{\mathrm{off}}$%
}  & %
\parbox[t]{4cm}{%
\raggedright$k_{\mathrm{on}}=\lambda_{\mathrm{on}}V\frac{V_{\mathrm{reac,A}}^{\mathrm{eff}}}{V-V_{\mathrm{ex,A}}}$
\linebreak{}
 $k_{\mathrm{off}}=\lambda_{\mathrm{off}}V\frac{V_{\mathrm{reac,B}}^{\mathrm{eff}}}{V-V_{\mathrm{ex,B}}}$%
}\tabularnewline
\bottomrule
\end{tabular}\caption{Summary of the iPRD-DB quantities for three different kinds of reversible
reactions: reversible association (see Sec.~\ref{sec:microscopic-distribution}),
reversible unimolecular conversion, and reversible bimolecular enzymatic
reaction (see Sec.~\ref{subsec:other-reactions}). Quantities are:
absolute proposal rates $\lambda$, proposal densities $q$, and acceptance
probabilities $\alpha=\min\{1,f\}$, as described in Sec.~\ref{sec:db-general}.
Superscript $+$ and $-$ denote the ``on'' and ``off'' process
respectively, corresponding to the definition of the reaction. $\mathbf{x}$
and $\mathbf{y}$ are the microscopic positions of particles. Constraints
describe for which microscopic parameters the acceptance probabilities
will be unity in the dilute limit.}
\label{tab:summary}
\end{table*}

\end{document}